%% file: normalDriven_main.tex
\documentclass{egpubl}
\usepackage{egsgp2021}
 
\SpecialIssuePaper         

\usepackage[T1]{fontenc}
\usepackage{dfadobe}  

\input{derekPre.tex}

\biberVersion
\BibtexOrBiblatex
\usepackage[backend=biber,bibstyle=EG,citestyle=alphabetic,backref=true]{biblatex} 
\electronicVersion 
\PrintedOrElectronic

\ifpdf \usepackage[pdftex]{graphicx} \pdfcompresslevel=9
\else \usepackage[dvips]{graphicx} \fi

\usepackage{egweblnk} 

\title{Normal-Driven Spherical Shape Analogies}

\author[Hsueh-Ti Derek Liu \& Alec Jacobson]
{
  {\parbox{\textwidth}{\centering Hsueh-Ti Derek Liu and Alec Jacobson \\ University of Toronoto}} 
}

\begin{document}
  
\teaser{
  \includegraphics[width=\linewidth]{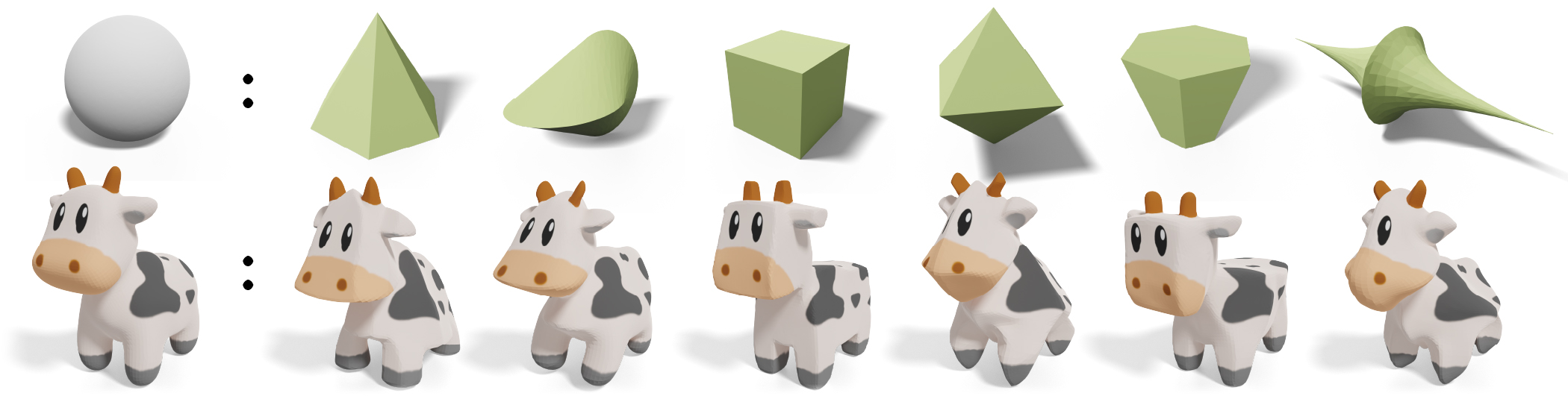}
  \caption{Our \emph{normal-driven spherical shape analogy} stylizes an input 3D shape (bottom left) by studying how the surface normal of a style shape (green) relates to the surface normal of a sphere (gray). }
  \label{fig:teaser}
}

\maketitle
\begin{abstract}
This paper introduces a new method to stylize 3D geometry. The key observation is that the surface normal is an effective instrument to capture different geometric styles. Centered around this observation, we cast stylization as a shape analogy problem, where the analogy relationship is defined on the surface normal. This formulation can deform a 3D shape into different styles within a single framework. One can plug-and-play different target styles by providing an exemplar shape or an energy-based style description (e.g., developable surfaces). Our surface stylization methodology enables Normal Captures as a geometric counterpart to material captures (MatCaps) used in rendering, and the prototypical concept of Spherical Shape Analogies as a geometric counterpart to image analogies in image processing. 
\end{abstract}  

\input{sections/introduction.tex}
\input{sections/related.tex}
\input{sections/method.tex}
\input{sections/analysis.tex}

\input{sections/applications.tex}
\input{sections/futureWork.tex}

\printbibliography   

\input{sections/appendix.tex}

\end{document}

%% file: derekPre.tex
\usepackage{wrapfig}
\usepackage{nicefrac}
\usepackage{mathtools}
\usepackage{booktabs} 
\usepackage{combelow} 
\usepackage[ruled,vlined,linesnumbered]{algorithm2e}
\usepackage{tabularx} 
\usepackage{colortbl} 

\usepackage{manfnt} 

\usepackage{amsmath}
\usepackage{amsfonts}
\usepackage{amsbsy}
\usepackage{empheq} 

\definecolor{derekBlue}{RGB}{144,210,236}
\definecolor{derekTableBlue}{RGB}{189,235,252}
\definecolor{iglGreen}{RGB}{153,203,67}
\definecolor{coralRed}{RGB}{250,114,104}
\definecolor{gray}{RGB}{180,180,180}
\definecolor{orange}{RGB}{255,69,0}

\SetCommentSty{commentFont}
\SetNlSty{textbf}{}{.}

\newcommand{\update}[1]{#1}

\newcommand{\refequ}[1] {Eq.~\ref{equ:#1}}

\newcommand{\reffig}[1] {Fig.~\ref{fig:#1}}
\newcommand{\reffignum}[1] {\ref{fig:#1}}

\newcommand{\refsec}[1] {Sec.~\ref{sec:#1}}

\newcommand{\refapp}[1] {App.~\ref{app:#1}}
\newcommand{\refappnum}[1] {\ref{app:#1}}
\newcommand{\refalg}[1] {Alg.~\ref{alg:#1}}

\DeclareMathOperator{\Tr}{Tr}
\DeclareMathOperator*{\argmax}{arg\,max}
\DeclareMathOperator*{\argmin}{arg\,min}

\newcommand{\R}{\mR}
\newcommand{\V}{\mV}

\newcommand{\dV}{\ve}
\newcommand{\DV}{\mE}
\newcommand{\U}{\mV'} 
\renewcommand{\u}{\vv'}
\newcommand{\dU}{{\ve'}}
\newcommand{\DU}{{\mE'}}

\newcommand{\n}{\hat{\vn}}
\newcommand{\nn}{\hat{n}}

\newcommand{\T}{T}
\newcommand{\tt}{\vt}
\newcommand{\sphere}{A}
\newcommand{\style}{A'}
\newcommand{\Vin}{B}
\newcommand{\Vout}{B'}
\newcommand{\Nsphere}{\widetilde{N}_{A'}}

\def\arap{{\textsc{arap}}\xspace}
\def\farap{{\textsc{farap}}\xspace}
\def\acap{{\textsc{acap}}\xspace}

\newcommand{\vecFont}[1]{\mathsf{#1}}

\def\ve{{\vecFont{e}}}

\def\vn{{\vecFont{n}}}

\def\vs{{\vecFont{s}}}
\def\vt{{\vecFont{t}}}
\def\vu{{\vecFont{u}}}
\def\vv{{\vecFont{v}}}

\newcommand{\matFont}[1]{\mathsf{#1}}
\def\mA{{\matFont{A}}}

\def\mE{{\matFont{E}}}
\def\mF{{\matFont{F}}}

\def\mK{{\matFont{K}}}

\def\mM{{\matFont{M}}}

\def\mQ{{\matFont{Q}}}
\def\mR{{\matFont{R}}}

\def\mV{{\matFont{V}}}
\def\mW{{\matFont{W}}}
\def\mX{{\matFont{X}}}

%% file: sections/introduction.tex
\section{Introduction}\label{sec:intro}



Analogies of the form $A:A'\dblcolon B:B'$ is a reasoning process that conveys \emph{$A$ is to $A'$ as $B$ is to $B'$}. This formulation has become a core technique for creating artistic 2D digital content, such as image analogies \cite{HertzmannJOCS01} in Photoshop \cite{photoshop} for image stylization and the Lit Sphere \cite{SloanMGG01} (a.k.a. MatCap) in ZBrush \cite{zbrush} for non-photorealistic renderings.
However, leveraging analogies to stylize 3D geometry is still at a preliminary stage because defining the analogy relationship on surface meshes requires dealing with irregular discretizations, curved metrics, and different topologies. 


In this paper, we introduce a step towards a more general 3D shape analogies, named \emph{spherical shape analogies}. We consider a specific case where $A$ is a unit sphere. This restriction enables us to operate on an input mesh $B$ with arbitrary topologies, boundaries, and geometric complexity. 
While not fully general, because $A$ is restricted to be a sphere, we demonstrate that this formulation can immediately achieve different geometric styles within a single framework. In \reffig{teaser}, we show that by providing different target style shapes $A'$ to the algorithm, we can turn the input shape $B$ into different styles.
In addition to stylization, our method can encompass many existing applications, such as developable surface approximation and PolyCube deformation. 

One key observation in our spherical shape analogies is that the surface normal is an effective instrument to capture geometric styles. Thus, we define the analogy relationship based on normals: we optimize a stylized shape $B'$ such that the relationship between the surface normals of $B$ and $B'$ is the same as the relationship between the surface normals of $A$ and $A'$
%

We realize this by casting it as a simple and effective normal-driven shape optimization problem which aims at deforming the input shape towards a set of desired normals. However, such an optimization problem is often difficult due to the nonlinearity of unit normals. We draw inspiration from previous works and apply a change of variables to accelerate the computation: instead of directly optimizing the vertex positions, we optimize a set of rotations that rotate the normals of the input mesh to the set of desired normals. 
Our simple formulation with the change of variables results in a generic stylization algorithm that runs at interactive rates.   

%% file: sections/related.tex
\begin{figure}
  \centering
  \includegraphics[width=3.33in]{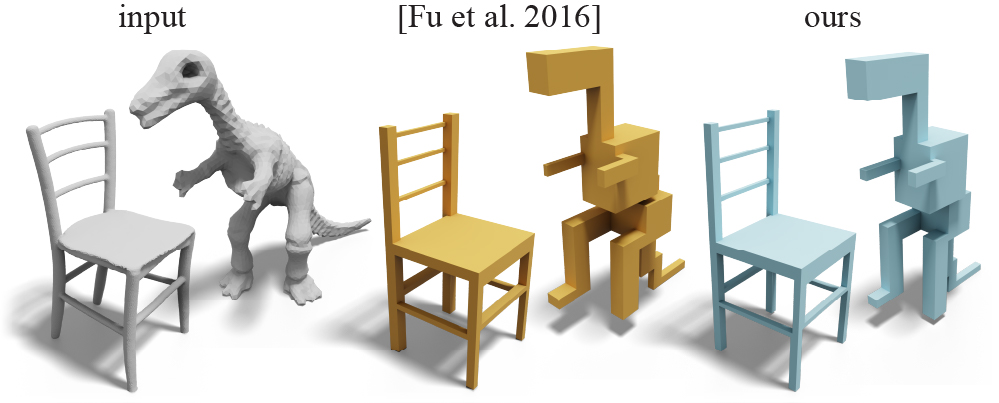}
  \vspace{-5pt}
  \caption{Our method can be used to create PolyCube shapes (blue) and obtain comparable results to \cite{FuBL16} (yellow).}  
  \label{fig:PolyCube_comparison}
  \vspace{-10pt}
\end{figure}
\begin{figure}
  \centering
  \includegraphics[width=3.33in]{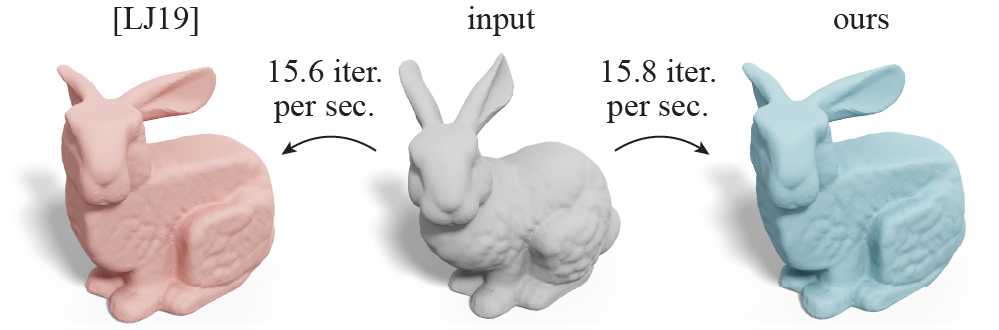}
  \vspace{-5pt}
  \caption{Although being more general for creating different geometric styles (e.g., \reffig{teaser}), our normal driven editing can also be applied to cubic stylization \cite{LiuCubic19}, achieving comparable performance (blue) to the previous method (red).} 
  \label{fig:cubic_comparison}
  \vspace{-10pt}
\end{figure}

\section{Related Work}
Our work shares similar motivations to computer-assisted image stylization pioneered by Haeberli \cite{Haeberli90}. But since our outputs are stylized 3D geometries, we focus the discussion on geometric stylization and geometric deformation methods. 

\subsubsection*{Analogy-based Geometric Stylization}
Many generative models have been proposed for creating stylized 3D objects, such as collage art \cite{gal20073d, Theobalt2007Collage}, manga style \cite{ShenLHC12}, cubic style \cite{LiuCubic19}, and neuronal homunculus \cite{ReinertRS12}. However, these methods are tailor-made for only a specific style.

\emph{Analogy} $\sphere:\style\dblcolon \Vin:\Vout$ is a powerful idea to achieve different stylization results within a single framework. 
This idea has inspired several design tools for images \cite{HertzmannJOCS01}, non-photorealistic renderings \cite{SloanMGG01,Shechtman16}, and curves \cite{HertzmannOCS02}. 
\update{Beyond 2D data, the idea of analogy has also been used for transferring 3D geometric details from one shape to another. 
We omit the discussion on methods that are not based on analogies, such as mesh cloning \cite{ZhouHWTDGS06, TakayamaSSIBS11} and geometric learning \cite{LiuKCAJ20, HertzHGC20, WangAKCS20, chen2021decor, li_cvpr21}, and focus on analogy-based techniques.}
Ma et al. \cite{MaHSKW14} propose a method for 3D style transfer based on patch-based assembly. However, their method cannot handle free-form deformations and requires the source and the exemplar shape to share a similar structure in order to compute high-quality correspondences. 
\update{Bhat et al. \cite{BhatIT04} propose a voxel-based texture synthesis method for transferring geometric details encoded in the volumetric grid. Berkiten et al. \cite{BerkitenHSMLR17} use metric learning for details represented as displacement maps. These methods are designed for high-frequency details (e.g., wrinkles on the surface).} In contrast, our spherical shape analogies focuses on larger scale free-form deformations. 
Albeit limited --- in our analogies $\sphere$ is restricted to the unit sphere --- our method enables a first step in this exciting direction.

\subsubsection*{Surface Normals in Shape Deformation}
A key insight of our spherical shape analogies is to leverage surface normals to capture geometric styles.
The surface normal is a fundamental geometric quantity and is ubiquitous in geometry processing.
A representative example is in the \emph{PolyCube} deformation \cite{TariniHCM04} where the goal is to optimize surface normals to be axis-aligned.
Gregson et al. \cite{GregsonSZ11} and Zhao et al. \cite{ZhaoLLZXG17} use the closest rotation from the surface normal to an axis-aligned direction to drive the PolyCube deformation. Huang et al. \cite{HuangJSTBD14} and Fu et al. \cite{FuBL16} propose to minimize energies defined on normals to create PolyCube shapes.
In architectural geometry design, surface normals are a main ingredient to characterize polygon meshes with planar faces. The methods proposed by Deng et al. \cite{DengPW11} and Poranne et al. \cite{PoranneOG13} utilize normals to formulate a \emph{distance-from-plane} constraint to encourage planarity. Tang et al. \cite{TangSGWP14} use the dot product between a face normal and its adjacent edge vectors to determine whether the vertices of a polygon are coplanar. 
\begin{figure}
  \centering
  \includegraphics[width=3.33in]{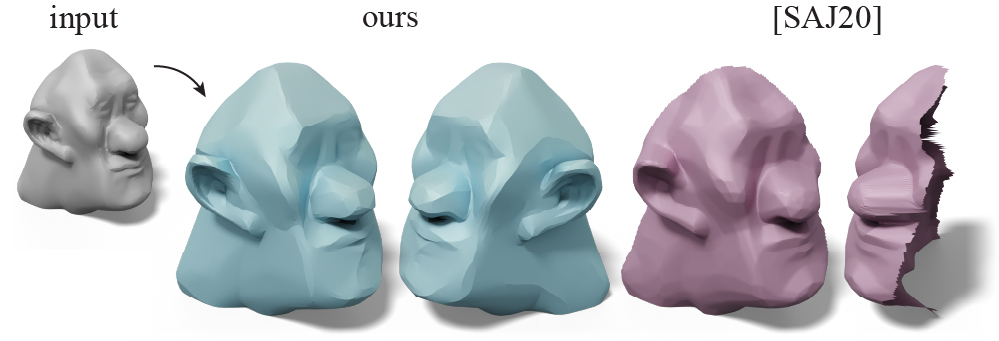}
  \vspace{-5pt}
  \caption{Sell{\'{a}}n et al. \cite{Sellan2020} propose a technique to make 2D heightfields developable (purple). In contrast, our method can create developable approximations for surface meshes in 3D (blue). } 
  \label{fig:silvia}
  \vspace{-10pt}
\end{figure}
\begin{figure}
  \centering
  \includegraphics[width=3.33in]{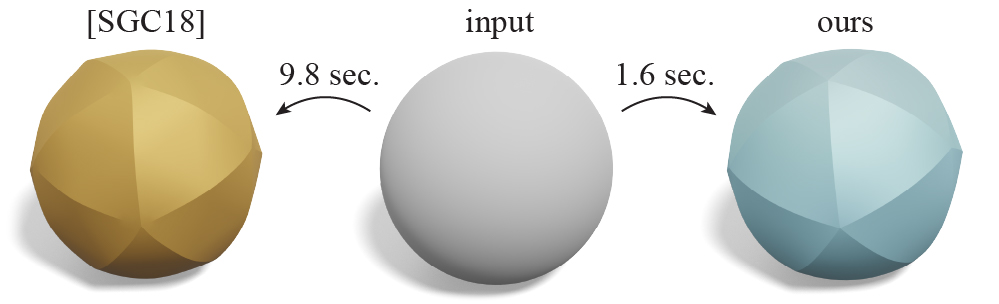}
  \vspace{-5pt}
  \caption{Compared to the method proposed by Stein et al. \cite{SteinGC18} for creating developable approximations (left), our method can create visually comparable results (right) with significant speed-ups.} 
  \label{fig:developable_comaprison}
  \vspace{-10pt}
\end{figure}
Characterizing whether a mesh can be flattened to 2D without stretching or shearing, a.k.a. \emph{developability}, also relies on surface normals. Stein et al. \cite{SteinGC18} characterize discrete developability based on the 1-ring face normals, and propose an algorithm to compute piecewise developable surfaces. Sell{\'{a}}n et al. \cite{Sellan2020} reformulate the developable energy into a convex semidefinite program for finding piecewise developable heightfields.
In addition to these examples, deforming shapes into the cubic style \cite{LiuCubic19, Fumero0R20}, constructing shape abstractions \cite{alexa2021polycover}, surface parameterization \cite{ZhaoSLZYLWGG20}, \update{and interactive mesh editing \cite{YuZXSBGS04, SorkineCLARS04}} are all related to surface normals. Many more examples can be found in the design of geometric filters, such as the Guided filter \cite{ZhangDZBL15}, the Shock filter \cite{PradaK15}, the Bilateral normal filter \cite{ZhengFAT11}, and the Total Variation mesh denoising \cite{ZhangWZD15}.

Our method can be adapted to these normal-based deformations. Compared to the PolyCube method \cite{FuBL16}, we achieve comparable quality (see \reffig{PolyCube_comparison}), but we can further generalize to polytopes (see \reffig{generalized_PolyCube}).
%
%
Compared to \cite{LiuCubic19} in cubic stylization (see \reffig{cubic_comparison}), we can achieve similar performance, but we can further generalize to many styles other than the cubic style (see \reffig{teaser}).
%
%
In developable surface approximation, in contrast to the method by Sell{\'{a}}n et al. \cite{Sellan2020}, our method can be applied to surface triangle meshes (see \reffig{silvia}) and is significantly faster than the method by Stein et al. \cite{SteinGC18} (see \reffig{developable_comaprison}).
%

\subsubsection*{Shape Deformation}
Our geometric stylization method can also be perceived as \update{a type of shape deformation method}. We share technical similarities with methods that deform a shape while addressing given modeling constraints.  
A common choice is to minimize the \emph{as-rigid-as-possible} (\arap) energy \cite{SorkineA07,IgarashiMH05, ChaoPSS10} while satisfying the constraints. This \arap energy measures the rigidity of local surface patches and favors detail-preserving smooth deformations.
In the case where locally rigid deformations are too constrained, the conformal energy \cite{CranePS11, VaxmanMW15} which preserves angles is commonly used.  In contrast to \arap, the conformal energy often triggers larger deformations as it allows both local uniform scaling and rigid transformations. 
In addition to mesh deformations, similar energies have also been used for parameterization \cite{LiuZXGG08}, shape optimization \cite{BouazizDSWP12}, and simulating mass-spring systems \cite{LiuBOK13}.
The \arap and conformal energies are also commonly used as regularization terms in mesh optimization problems, such as reconstruction \cite{IzadiF14}, surface registration \cite{HuangAWG08, YoshiyasuMYK14}, PolyCube construction \cite{HuangJSTBD14}, and surface stylization \cite{LiuCubic19}. 
Their popularity comes from the property that they favor smooth deformations and are amenable to fast optimizations.
For the same reasons, we also use these as our regularization energies for interactive modeling tasks (see \reffig{regularizations}). 

%% file: sections/method.tex
\section{Spherical Shape Analogies}
\begin{figure}
  \centering
  \includegraphics[width=3.33in]{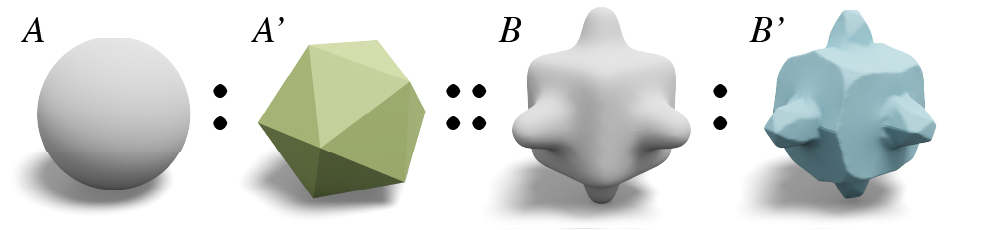}
  \vspace{-5pt}
  \caption{We generate an output shape $\Vout$ that relates to the input $\Vin$ in the same way as how the surface normal of a given primitive $\style$ relates to the surface normal of a sphere $\sphere$. } 
  \label{fig:analogies}
  \vspace{-10pt}
\end{figure}
Our main idea is to use surface normals to capture the style of 3D objects: if two shapes share a similar normal ``profile'', we consider them to exhibit the same geometric style. 
Centered around this observation, as discussed in \refsec{intro}, we propose an analogy-based stylization method to translate the relationship between the normals of $\style, \sphere$ to create a stylized output shape $\Vout$ (see \reffig{analogies}). 
Throughout the paper, we use green color to denote the target style shape $\style$, gray color to denote the input shape $\Vin$, and blue color to denote the output stylized shape $\Vout$.

\begin{figure}
  \centering
  \includegraphics[width=3.33in]{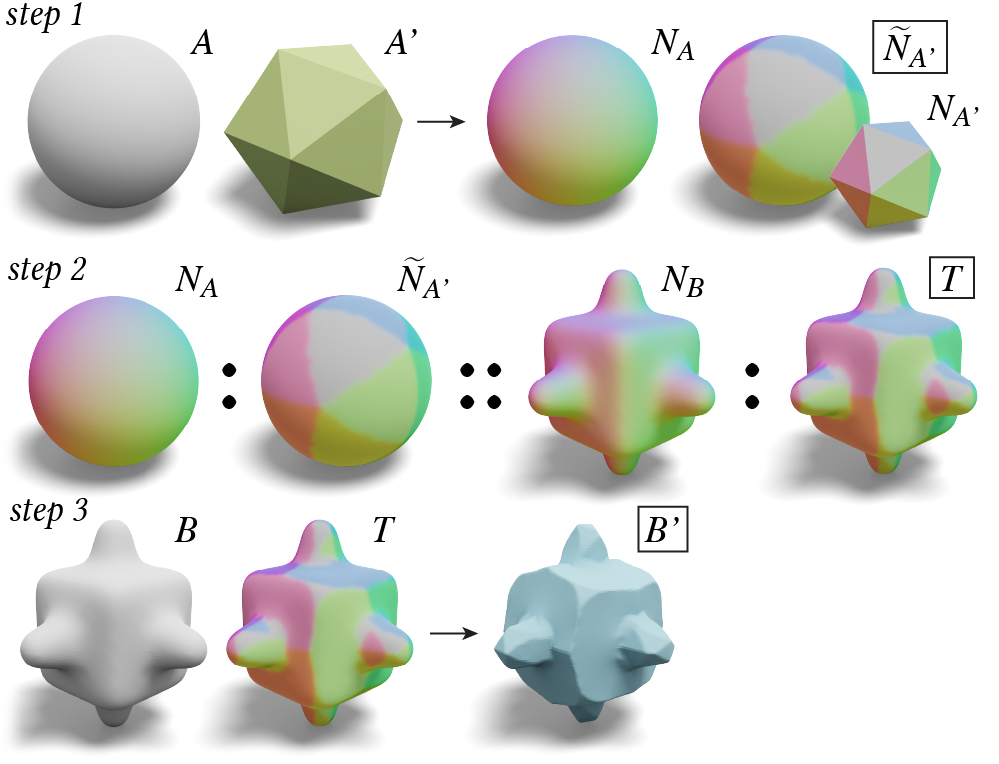}
  \vspace{-5pt}
  \caption{Our algorithm defines the analogous relation based on the surface normals. We first map the normals of the style shape $N_{\style}$ to a unit sphere to obtain $\Nsphere$ (top row), transfer the relationship between $N_\sphere$ and $\Nsphere$ to the input shape to obtain the target normals $\T$ (middle row), then optimize the input shape $\Vin$ so that the actual output normals are aligned with the target normal $\T$ (bottom row). } 
  \label{fig:analogies_gauss}
  \vspace{-10pt}
\end{figure}

Our algorithm consists of three simple steps, described in \reffig{analogies_gauss}: (1) we map the surface normal of $\style$ to a unit sphere $\sphere$ in order to compute target normals on a sphere $\Nsphere$, (2) we construct analogous target normals $\T$ that relate to $N_{\Vin}$ the same way $\Nsphere$ relate to $N_{\sphere}$, (3) we take $\Vin, \T$ as inputs and generate the stylized shape $\Vout$ whose normals approximate $\T$ via optimization. 

\subsection{Generating $\Nsphere$}\label{sec:generateNS}
Depending on the provided style shape $\style$ or user preferences, we consider three ways to get a set of target normals on a sphere $\Nsphere$.

\textit{1. Closest normals.}\ 
The simplest case is when the style shape $\style$ is a simple convex shape with only few distinct face normals (e.g., icosahedron). We compute $\Nsphere$ simply via snapping the normals of the sphere $N_{\sphere}$ to the nearest normal in the style shape $N_{\style}$.

\textit{2. Spherical parameterization.}\ 
For a generic genus-0 shape (e.g., smooth or concave), we compute its parameterization to a sphere using, for example, conformalized mean curvature flow \cite{KazhdanSB12}. Then $\Nsphere$ can be computed from the spherical parameterization.
\begin{figure}
  \centering
  \includegraphics[width=3.33in]{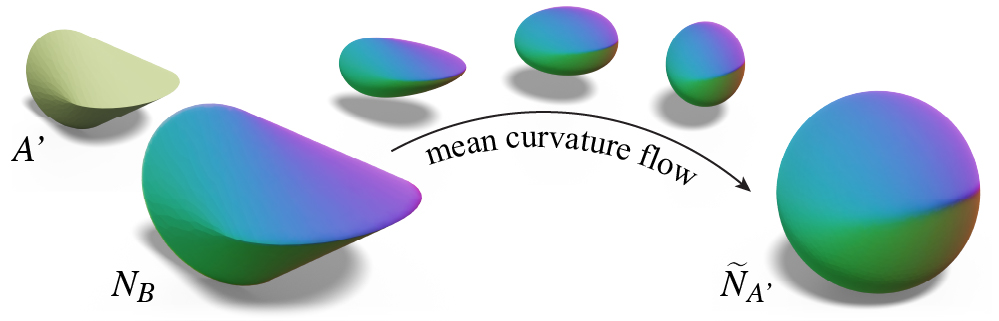}
  \vspace{-5pt}
  \caption{Given a style shape $\style$, we run mean curvature flow \cite{KazhdanSB12} to map the normals of style shape $N_{\style}$ to a sphere as $\Nsphere$.} 
  \label{fig:meanCurvatureFlow}
  \vspace{-10pt}
\end{figure}

\textit{3. Normal Capture.}\ 
If one desires more control, one can manually specify $\Nsphere$, in the spirit of how \emph{MatCap} (material capture \cite{SloanMGG01}) is used in rendering. We can then skip the first step in \reffig{analogies_gauss} and move on to the second step using the user-provided $\Nsphere$.

\subsection{Generating $\T$}\label{sec:generateT}
\begin{wrapfigure}[6]{r}{1.33in}
	\raggedleft
  \vspace{-12pt}
	\hspace*{-0.7\columnsep}
	\includegraphics[width=1.48in, trim={6mm 0mm -1mm 0mm}]{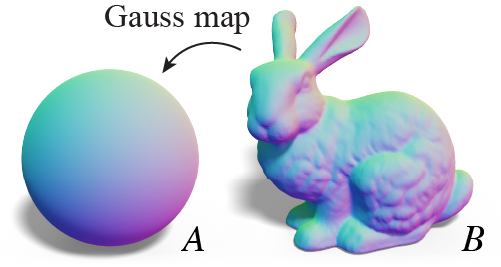}
	\label{fig:gaussmap}
\end{wrapfigure}
Generating target normals $\T$ on the input shape $\Vin$ using analogy requires the correspondences between $\sphere$, $\Vin$. We compute the map using the \emph{Gauss map}, leveraging the fact that our $\sphere$ 
is always a unit sphere (see the inset, where we use colors to visualize the correspondences).
\update{Specifically, the unit normal vector of each element (e.g., vertex or face) on the input shape $\Vin$ can be equivalently interpreted as a point on the unit sphere $\sphere$. Thus, we can easily map signals from $\sphere$ back to $\Vin$.}
Once the correspondences are obtained via the normals of input shape $N_{\Vin}$, we can trivially compute $\T$ by ``pasting'' $\Nsphere$ on top of $\Vin$.

\subsection{Generating $\Vout$}
\begin{figure}
  \centering
  \includegraphics[width=3.33in]{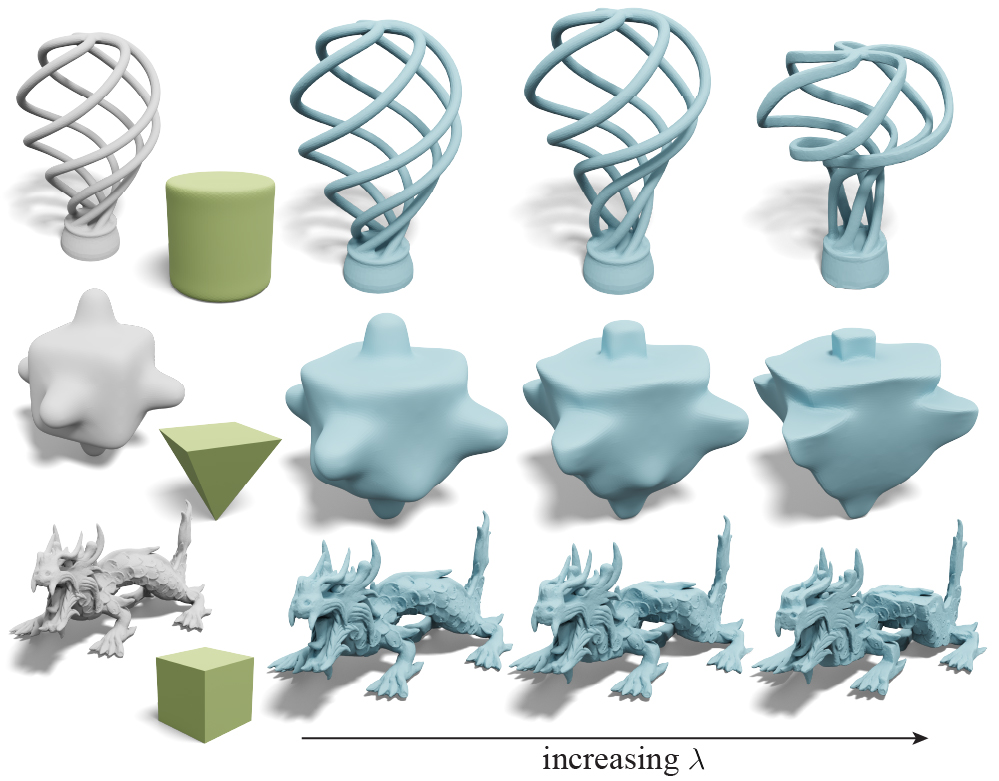}
  \vspace{-5pt}
  \caption{\update{The $\lambda$ parameter in \refequ{arapNormalDriven} controls the balance between preserving the original shape and satisfying the desired style. We show different stylization results with increasing $\lambda$. \copyright Spiral Light Bulb (top) by benglish under CC BY-SA.}} 
  \label{fig:lambda}
  \vspace{-10pt}
\end{figure}

After obtaining a set of target normals $\T = \{ \tt_k \}$ for each vertex $k$, our goal is to obtain a deformed output shape $\Vout$ whose surface normals approximate $\T$. 
Let $\V$ be a matrix of vertex locations with size $|\V|$-by-3 and $\mF$ be the face list with size $|\mF|$-by-3 of the input shape $\Vin$. Our output shape $\Vout$ is a deformed version of the input shape and we use $\U$ to denote the $|\V|$-by-3 matrix of the deformed vertex locations. We formulate the normal-driven deformation as an energy optimization in the following form:
\begin{align}
  \min_{\U} \sum_{k \in \V} E_R(\vv_k,\vu_k) + \lambda a_k \| \nn_k(\U) - \tt_k \|_2^2,
\end{align} 
where $E_R$ denotes a regularization energy to preserve the details of the input mesh, and the second part measures the squared distance from the output unit surface normal $\n_k(\U)$ to the target output normal $\tt_k$ at vertex $k$. We use $a_k$ to denote the Voronoi area of the vertex $k$, $\lambda$ is a weighting parameter to control the balance between the two terms, and $\vv_k$ ($\vu_k$) is the input (output) location of vertex $k$.
In \reffig{lambda}, we can observe that using a small $\lambda$, the method preserves the input shape $\Vin$. Using a bigger $\lambda$, the method favors in deforming the shape more into the style of $\style$.
The choice of $E_R$ depends on the user's intent. One can apply different regularizations to obtain different results.
For the purposes of this exposition, we introduce our optimization based on \arap regularization in \refsec{normalDriven_arap}. We discuss how to extend to other regularizations in \refsec{regularizations}. 
%

\subsection{Normal-Driven Optimization with \arap}\label{sec:normalDriven_arap}
We use $\dV_{ij} \coloneqq \vv_j - \vv_i \in \mathbb{R}^3$ to denote the edge vector between vertices $i, j$ on the original mesh, and $\dU_{ij} \coloneqq \u_j - \u_i$ for the edge vectors on the deformed mesh. We can write down the energy that uses \arap regularization as
\begin{align}\label{equ:arapOutputNormal}
  \min_{\U,\R}\ \sum_{k \in  V} \underbrace{\sum_{i,j \in \mathcal{N}_k} w_{ij} \| \R_k \dV_{ij} -  \dU_{ij} \|^2_2}_{E_\arap} + \lambda a_k \| \nn_k(\U) - \tt_k \|_2^2,
\end{align}

\begin{wrapfigure}[6]{r}{0.9in}
	\raggedleft
	\vspace{-14pt} 
	\hspace*{-0.9\columnsep}
	\includegraphics[width=1.25in, trim={5mm 0mm 1mm 0mm}]{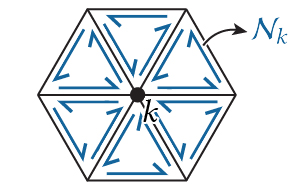} 
\end{wrapfigure}
We use $\mathcal{N}_k$ to denote the edge vectors of the \emph{spokes and rims} at vertex $k$ (see the inset) \cite{ChaoPSS10}, $\R_k \in \text{SO(3)}$ to denote a 3-by-3 rotation matrix defined on $k$, and $w_{ij}$ is the cotangent weight of edge $i,j$ \cite{PinkallP93}. 
However, this energy is difficult to optimize because the term $\nn_k(\U)$ is non-linear in $\U$. 

\begin{figure}
  \centering
  \includegraphics[width=3.33in]{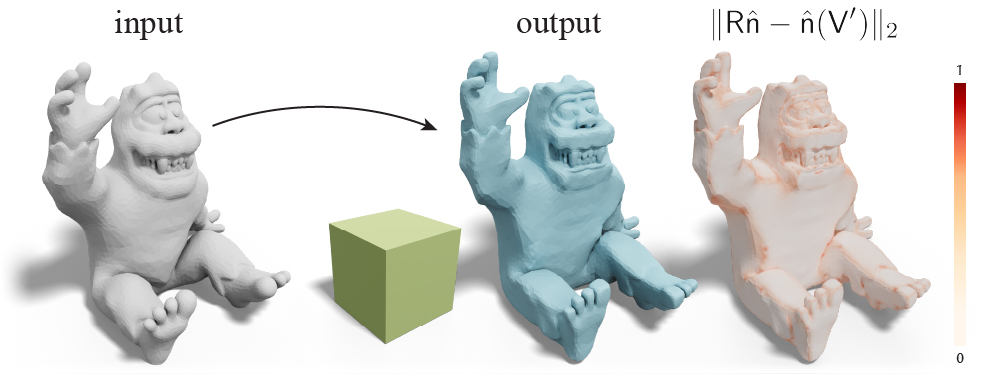}
  \vspace{-5pt}
  \caption{Empirically, we show that rotated input normals $\R_k \n_k$ is a good approximation of the area-weighted output vertex normals $\nn_k(\U)$. We can observe that the error mostly occurs on the high-curvature regions (right). \copyright Proto Paradigm under CC BY.} 
  \label{fig:RN_difference}
  \vspace{-10pt}
\end{figure}

We adapt the observation made in \cite{LiuCubic19} that the space of unit vectors can be captured by rotations. Thus, we can perform a change of variables by replacing $\nn_k(\U)$ with the \emph{rotated} unit normal of the input mesh $\R_k \n_k$ as
\begin{align} \label{equ:arapNormalDriven}
  \boxed{
  \min_{\U,\R}\ \sum_{k \in V} \sum_{i,j \in \mathcal{N}_k} w_{ij} \| \R_k \dV_{ij} -  \dU_{ij} \|^2_2  + \lambda a_k \| \R_k \n_k  - \tt_k \|_2^2,}
\end{align}
where $\n_k$ is the $k$th unit vertex normal of the input mesh computed via area-weighted average of face normals, which is constant throughout the optimization. This $\R_k \n_k$ can be perceived as an approximation of the area-weighted vertex normals of the output mesh $\nn_k(\U)$. In \reffig{RN_difference}, we visualize the difference between the output normals $\nn_k(\U)$ and the rotated input normals $\R_k \n_k$. We can notice that $\R_k \n_k$ is a decent approximation of the output vertex normals computed via area-weighted average. 
\update{We can observe that error tend to concentrate on high-curvature regions because discrete vertex normals are less accurate along on those regions and the \arap regularization encourages smooth deformation.}
This change of variables allows us to solve for $\R_k$s in parallel and make this energy quadratic in $\U$. In addition, the fact that $\R_k$ is shared across the \arap term and the normal term enables us to jointly consider both the regularization and the normal terms when obtaining the deformed vertex locations $\U$.

We minimize this energy via the local/global strategy \cite{SorkineA07}, where the local step involves solving a set of small \emph{Orthogonal Procrustes} problems and the global step amounts to a linear solve. For the sake of reproducibility, we reiterate the local-global steps for our energy in \refapp{localStep}, \refappnum{globalStep}. 
%
%
Non-linear methods, such as Newton’s method, could be applied to our scenario. It is however far slower than the local-global optimization since a single iteration of the Newton's method could be more expensive than 100 iterations of the local-global iterations (see \cite{LiuBK17}). Thus, it is less suitable for our interactive applications. 
Further accelerating our solver using other optimization methods (e.g., \cite{KovalskyGL16, PengDZGQL18, ZhuBK18}) should be possible, but is left as future work.

%% file: sections/analysis.tex
\section{Extensions \& Analysis}\label{sec:analysis} 
In this section, we introduce its extensions to different regularizations and how to handle cases where target normals $\T(\Vout)$ are a function of output geometry.   

\subsection{Different Regularizations}\label{sec:regularizations}
In addition to $E_\arap$, the normal-driven optimization supports different regularization energies for different modeling intents. 
One could use \arap when the goal is to produce a smooth deformation that preserves surface details.
If one wants to produce a non-smooth deformation (e.g., sharp creases) while preserving local rigidity, one could instead use a \emph{face-only} \arap energy $E_\farap$ \update{discussed in \cite{ZhaoG16a, LeviG15}} which consists of only the membrane term.
If one is interested in preserving the textures and allowing local scaling, one could use an \emph{as-conformal-as-possible} energy $E_\acap$ \cite{BouazizDSWP12}.

\begin{wrapfigure}[4]{r}{0.53in}
	\raggedleft 
	\vspace{-16pt}
	\hspace*{-0.9\columnsep}
	\includegraphics[width=0.85in, trim={3mm 0mm 1mm 0mm}]{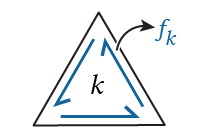} 
	\label{fig:stretchElement} 
\end{wrapfigure} 
\textit{Face-only \arap.}\ 
The core idea is to remove the bending term from \arap and only measure the membrane term \cite{TerzopoulosPBF87}, so that two adjacent triangles can bend freely. We achieve this by applying the idea from \cite{ZhaoG16a, LeviG15} which only measures the \arap energy over the three edge vectors of a face $f_k$ (see the inset), instead of the spokes and rims $\mathcal{N}_k$. Precisely, we can write this ``face-only'' \arap regularization $E_\farap$ as
\begin{align}\label{equ:farap}
  E_\farap(\U, \R) =  \sum_{k \in F} \sum_{i,j \in f_k} w_{ij} \| \R_k \dV_{ij} -  \dU_{ij} \|^2_2,
\end{align}

\textit{As-conformal-as-possible.}\ 
If the goal is to create novel geometric details, it is crucial to allow non-rigid deformations. However, an arbitrary deformation may lead to undesirable behaviors, such as badly shaped triangles. Thus constraining the angle preservation, a.k.a. conformality, will be a suitable regularization.
Specifically, we use the \acap energy $E_\acap$ in \cite{BouazizDSWP12} as our regularization 
\begin{align}\label{equ:acap}
  E_\acap(\U, \R, \vs) =  \sum_{k \in F} \sum_{i,j \in \mathcal{N}_k} w_{ij} \| s_k \R_k \dV_{ij} -  \dU_{ij} \|^2_2,
\end{align}
where $s_k$ is a scalar representing the scaling of local patch. One can compute the optimal $s_k$ analytically via the method by Sch{\"o}nemann et al. \cite{schonemann1970fitting} (see \refapp{regularizations}).

\begin{figure}
  \centering
  \includegraphics[width=3.33in]{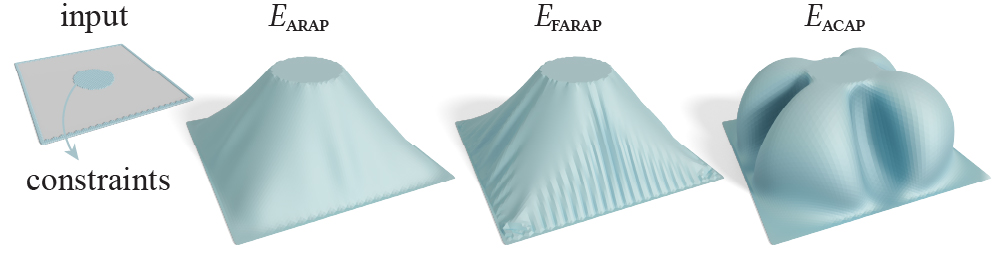}
  \vspace{-5pt}
  \caption{Different regularizations favor different behaviors. Given a sheet (gray), we pull up the center part (central blue dots) and shrink the boundary (blue dots on the boundary), then we minimize each regularization energy to determine the unconstrained vertices. We can observe that $E_\arap$ favors rigid and smooth interpolation, $E_\farap$ favors sharp bending between triangles, and $E_\acap$ favors to preserve angles while allowing local scaling. } 
  \label{fig:regularizations}
  \vspace{-10pt}
\end{figure}

Deploying these regularizations $E_\farap, E_\acap$ requires only a few changes in the optimization steps. Deploying $E_\farap$ only involves changing the incidence matrix. Deploying $E_\acap$ only requires adding one more line of code in the local step to solve an \emph{isotropic orthogonal Procrustes} problem \cite{schonemann1970fitting}. We detail such changes in \refapp{regularizations}. In \reffig{regularizations}, we apply the same deformation to a sheet but with different regularizations. We can perceive that different regularizations favor drastically different solutions.  

Our framework allows one to easily plug-and-play different regularizations. Specifically, we use $E_\arap$ for applications that favor smooth deformation (e.g., \reffig{primStyles}), $E_\farap$ for creating sharp creases (\reffig{generalized_PolyCube}, \reffignum{developable}), and $E_\acap$ when one wants to manipulate geometric details such as in \reffig{synthesis}.
\begin{figure}
  \centering
  \includegraphics[width=3.33in]{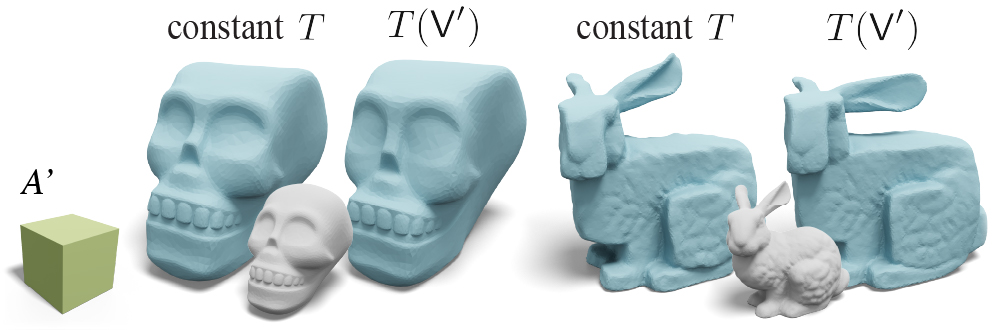}
  \vspace{-5pt}
  \caption{Setting the target normal $\T$ as a constant or treating it as a function of the output mesh $\T(\Vout)$ leads to different local minima. In many cases (left pair), both options lead to similar looking results. But setting $\T$ as a constant may result in an undesirable local minimum in some cases (right), such as the ears of the bunny.} 
  \label{fig:update_t}
  \vspace{-10pt}
\end{figure} 

\subsection{Dynamic Target Normals}
%
Our method converges to a local minimum. 
Empirically, we observe that treating the target normal $\T$ as a constant throughout the optimization may work fine perceptually in many cases (see the left pair in \reffig{update_t}). However, constant $\T$ may lead to an undesirable local minimum due to a sub-optimal assignment of $\T$ (see the right pair in \reffig{update_t}).
Inspired by Projective Dynamics \cite{BouazizMLKP14}, a simple solution to avoid such local minima is to treat $\T$ as a function of $\Vout$ ($N_{\Vout}$ specifically), and update $\T$ at every iteration.
%
%
We summarize the pseudo code in \refalg{normalDriven}. If $\T$ is a constant throughout the optimization, one can simply skip the optional step at line 9.
\IncMargin{1em}
\begin{algorithm}[t]
    \SetKwInOut{Input}{Input}
    \SetKwInOut{Output}{Output}
    \caption{Normal-driven optimization}
    \label{alg:normalDriven}
    \Indentp{-1em}
        \Input{$\,$A triangle mesh $\V, \mF$ and a weight $\lambda$}
        \Output{$\,$Deformed vertex positions $\U$}
        \BlankLine
    \Indentp{1em}
        compute $\Nsphere$ \hfill \text{\color{iglGreen} // step 1, Sec. 3.1}\\
        compute $\T$ from $\nn(\V)$ \hfill \text{\color{iglGreen} // step 2, Sec. 3.2}\\
        $\n \leftarrow \nn(\V)$ \hfill \text{\color{iglGreen} // compute input surface normals}\\
        $\mQ, \mK \leftarrow precompute(\V,\mF)$ \hfill \text{\color{iglGreen} // see App. B}\\
        $\U \leftarrow \V$\\
        \While{\textit{not converge}}{
          $\mR \leftarrow local\text{-}step\,(\U, \n, \T, \lambda)$  \hfill \text{\color{iglGreen} // App. A}\\
          $\U \leftarrow global\text{-}step\,(\mR, \mQ,\mK)$ \hfill \text{\color{iglGreen} // App. B}\\
          compute $\T$ from $\nn(\U)$ \hfill \text{\color{iglGreen} // (optional) for dynamic $\T$}\\
        }
\end{algorithm} 
\DecMargin{1em}

\begin{wrapfigure}[7]{r}{1.33in}
	\raggedleft
  \vspace{-12pt}
	\hspace*{-0.7\columnsep}
	\includegraphics[width=1.55in, trim={6mm 0mm -1mm 0mm}]{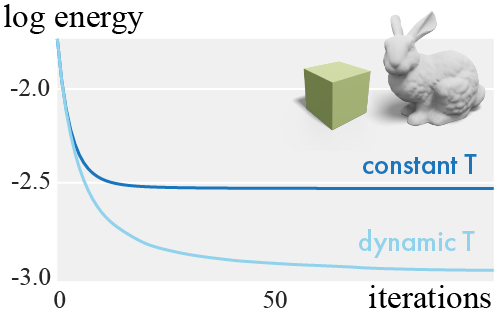}
	\label{fig:convergence}
\end{wrapfigure}
In terms of convergence, in the case where $\T$ is constant, the convergence behaves the same as the original \arap \cite{SorkineA07}, where the energy decreases monotonically. In the case where $\T$ is dependent to $\Vout$, we do not guarantee a monotonic decrease in energy, but the optimization still converges in our experiments. In the inset, we visualize the convergence plot for examples in \reffig{update_t}. 

\begin{wrapfigure}[7]{r}{1.33in}
	\raggedleft
  \vspace{-12pt}
	\hspace*{-0.7\columnsep}
	\includegraphics[width=1.55in, trim={6mm 0mm -1mm 0mm}]{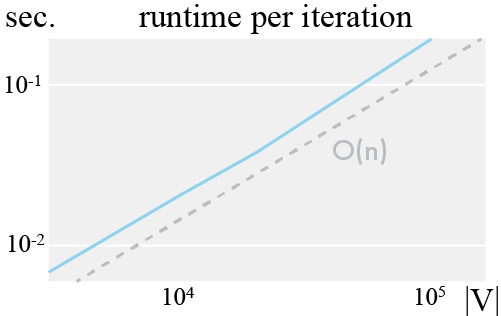}
	\label{fig:runtime}
\end{wrapfigure}
We implement our algorithm in C++ with Eigen \cite{eigenweb} and evaluate our method on a MacBook Pro with an Intel i5 2.3GHz processor.
Our method runs 24 iterations per second for a mesh with around 20k vertices. We report a complete picture of our runtime in the inset. The local step will be the computation bottleneck for meshes with less than 20k vertices, but further acceleration can be achieved via the method by Zhang et al. \cite{ZhangFast2021}. 
Typically, within the first 10 iterations, our method can achieve a visually similar result compared to the converged solution. This property enables us to build an interactive tool for users to play with different style shapes $\style$ or artistic controls.

%% file: sections/applications.tex
\begin{figure}
  \centering
  \includegraphics[width=3.33in]{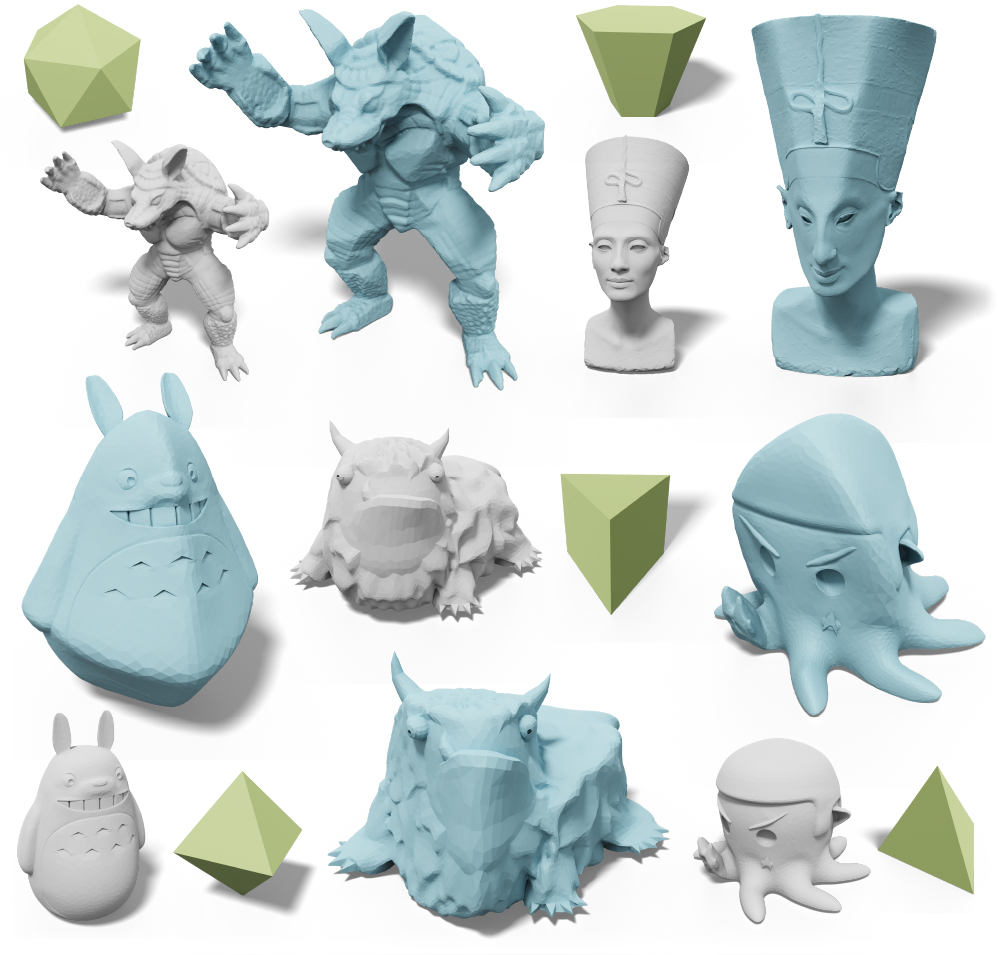}
  \vspace{-5pt}
  \caption{Given an input shape (gray), our approach can transfer the style of a primitive shape (green) to obtain a stylized output shape (blue). \copyright Johannes (bottom left), Joseph Larson (bottom middle), and Angelo Tartanian (bottom left) under CC BY. The Nefertiti mesh (top right) was scanned by Nora Al-Badri and Jan Nikolai Nelles from the Nefertiti bust.} 
  \label{fig:primStyles}
  \vspace{-2pt}
\end{figure}
\begin{figure}
  \centering
  \includegraphics[width=3.33in]{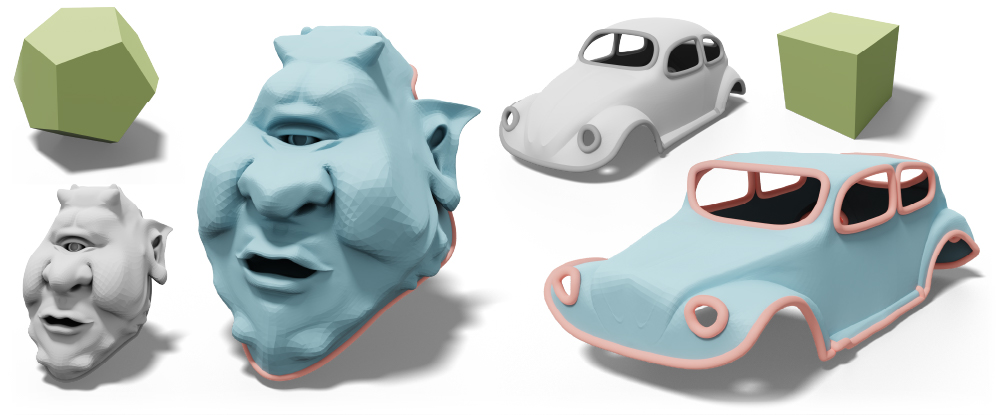}
  \vspace{-5pt}
  \caption{Even for input shapes with boundaries (gray), our method is still applicable to transfer the style of primitive shapes (green) to obtain the stylized output shape (blue).} 
  \label{fig:primStyles_bd}
  \vspace{-2pt}
\end{figure}
\begin{figure}
  \centering
  \includegraphics[width=3.33in]{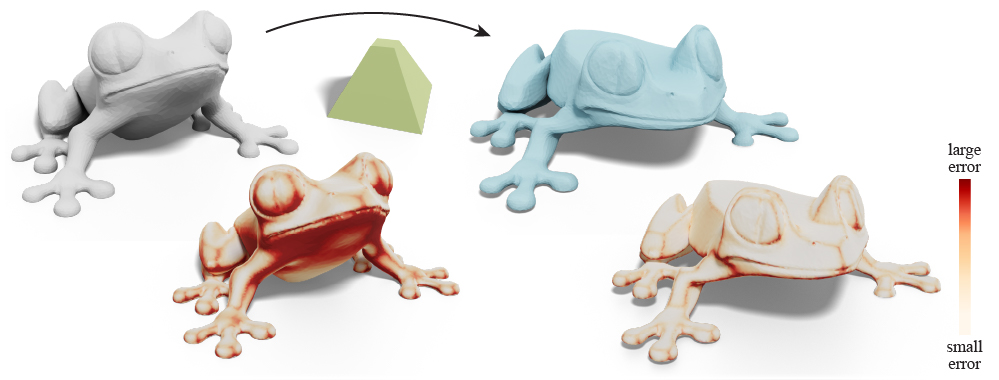}
  \vspace{-5pt}
  \caption{We visualize the difference between the mesh normals and the normals of the style shape. Our normal-driven optimization effectively reduce the difference to the target normals. \copyright Morena Protti under CC BY.} 
  \label{fig:quan_normal}
  \vspace{-5pt}
\end{figure}
\begin{figure}
  \centering
  \includegraphics[width=3.33in]{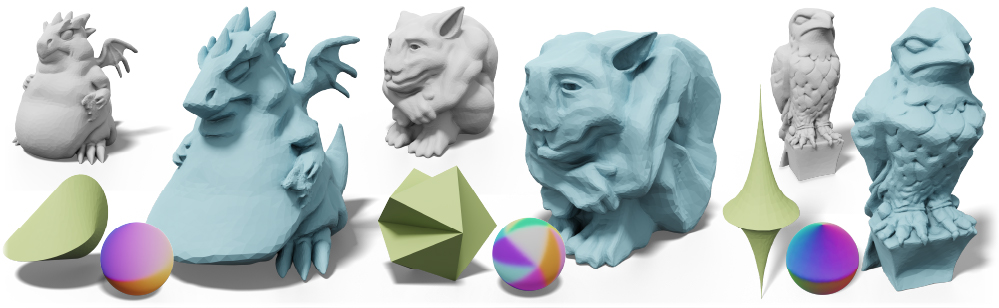}
  \vspace{-5pt}
  \caption{\update{When the input shape is} smooth or non-convex, we use the mean curvature flow (see \reffig{meanCurvatureFlow}) to obtain target normals to proceed the optimization. We deform the input shapes (gray) to exhibit the style of an oloid (left, green), a Jessen's icosahedron (middle, green), and a tractricoid (right, green), respectively. \copyright Splotchy Ink (left), fong182 (middle), and Colin Freeman (right) under CC BY.} 
  \label{fig:shape_gauss}
  \vspace{-10pt} 
\end{figure}
\begin{figure}
  \centering
  \includegraphics[width=3.33in]{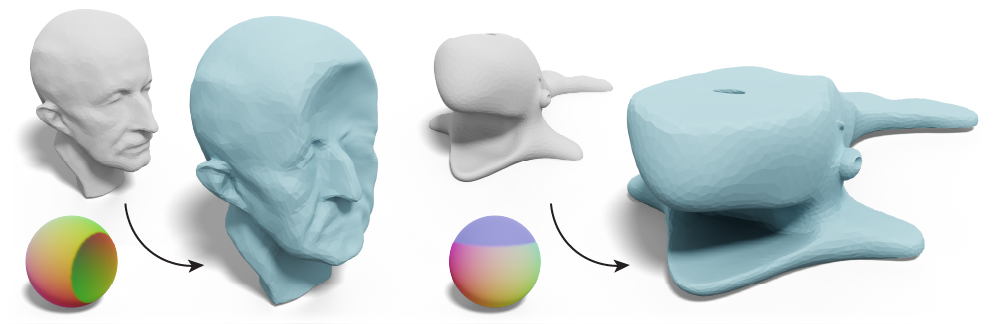}
  \vspace{-5pt}
  \caption{One can manually specify the target normals on a sphere (Normal Captures) for full control, and deform the input shape (gray) to the style (blue) prescribed by the colored sphere. \copyright MakerBot (right) under CC BY.} 
  \label{fig:manual_gauss}
  \vspace{-10pt} 
\end{figure}

\section{Applications}
%
The major benefit of our analogy-based stylization method is that one can plug-and-play different style shapes to obtain different results. \update{When one provides} convex primitives with few distinct face normals, we can simply use the method discussed in \refsec{generateNS} to turn an input shape into the style of the primitive (see \reffig{primStyles}, \reffignum{primStyles_bd}). In \reffig{quan_normal}, we also quantitatively show that our method can effectively reduce the difference between mesh normals and the normals of a primitive. 
If the provided style shape is smooth or non-convex, where the simple closest normal may fail to capture the style, one could use a spherical parameterization described in \refsec{generateNS} to achieve the stylization.
If desiring more user controls, one could ``paint'' the desired surface normals on a unit sphere (see \refsec{generateNS}), and then transfer the style of the painted normals directly to the input (see \reffig{manual_gauss})

\subsection{PolyCube Deformation}
\begin{figure}
  \centering
  \includegraphics[width=3.33in]{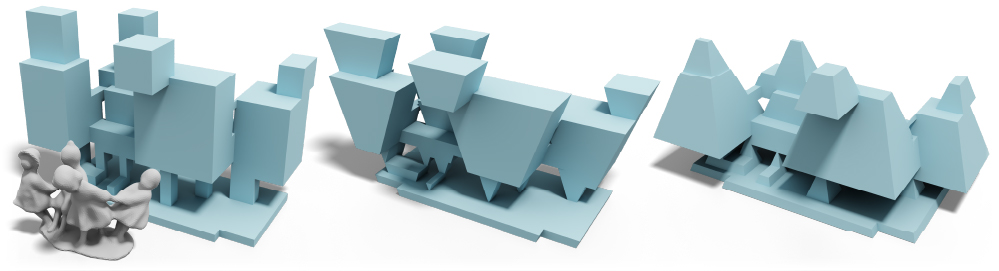}
  \vspace{-5pt}
  \caption{\update{By using} different sets of normals, we can generalize the PolyCube method (left) to create polygonal boxed maps.} 
  \label{fig:generalized_PolyCube}
  \vspace{-10pt} 
\end{figure}

If one is interested in PolyCube maps \cite{TariniHCM04}, we can adapt normal driven editing to create PolyCube maps,following the observation in \cite{ZhaoLLZXG17}. Specifically, we need to use a cube as a style shape and move the pre-computation step in \refalg{normalDriven} to the optimization loop. 
Moving the pre-computation in the loop would no longer preserve the original details, which is desirable for creating PolyCube shapes. This modification may also lead to badly shaped triangle. 
When these faces appear, a quick solution is to move the vertex towards the 1-ring average by a small amount to improve triangle quality.
For the sake of comparison, we use the same PolyCube segmentation as in \cite{FuBL16} and show that we can achieve comparable results in \reffig{PolyCube_comparison}. 
We can further generalize the PolyCube map to other polygonal boxes by specifying non-cube normals (see \reffig{generalized_PolyCube}).

\begin{figure}
  \centering
  \includegraphics[width=3.33in]{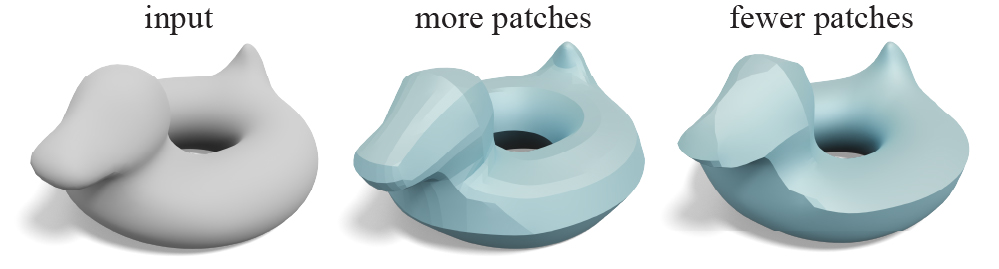}
  \vspace{-5pt}
  \caption{Stein et al. \cite{SteinGC18} control the patches on the developable surfaces via remeshing the input. We, instead, can control the amount of creases (middle, right) by tuning a single parameter (see \refapp{developable}).}
  \label{fig:control_developable}
  \vspace{-10pt} 
\end{figure}

\subsection{Developable Surface Approximation}
So far we have only considered an explicit shape or a set of painted normals as our style shape. Here we further extend our method to support an energy that describes a certain style. In particular, we consider the target normal $\T$ is computed via an optimization
\begin{align}
  \T = \argmin_\T f(\Vout),
\end{align}
and, similar to the case where $\T$ is dependent to $\Vout$, we update $\T$ at every iteration in the local/global solve. 

\begin{wrapfigure}[7]{r}{1.33in}
	\raggedleft
    \vspace{-14pt}
	\hspace*{-0.7\columnsep}
	\includegraphics[width=1.48in, trim={6mm 0mm -1mm 0mm}]{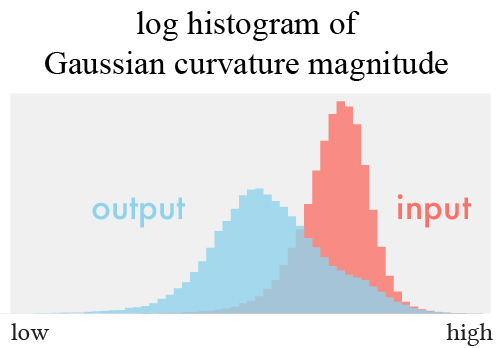}
	\label{fig:K_hist}
\end{wrapfigure}
\begin{figure}
  \centering
  \includegraphics[width=3.33in]{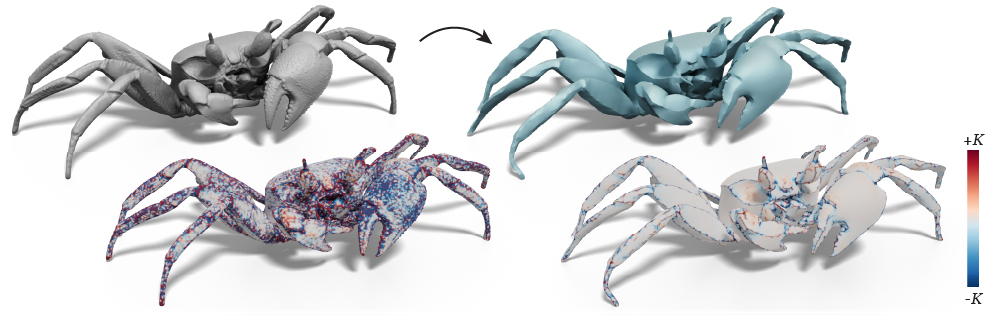}
  \vspace{-5pt}
  \caption{We use our normal driven editing to deform the input shape (gray) into a piece-wise developable approximation (blue). In the bottom row, we visualize the Gaussian curvature concentrates on the creases after the deformation, leading to a piece-wise developable shape. \copyright Oliver Laric under CC BY-NC-SA.} 
  \label{fig:K_developable}
  \vspace{-10pt}
\end{figure}

We evaluate this extension via setting $f$ to be the discrete developability energy proposed in \cite{SteinGC18}, with details provided in \refapp{developable}. Compared to the original method, our approach contains a regularization term in addition to the developable energy, thus our optimization requires no remeshing and results in the faster optimization (see \reffig{developable_comaprison}).  
In \reffig{control_developable}, we further show that our framework enables one to control the number of creases in the results.
%
%
\begin{figure}
  \centering
  \includegraphics[width=3.33in]{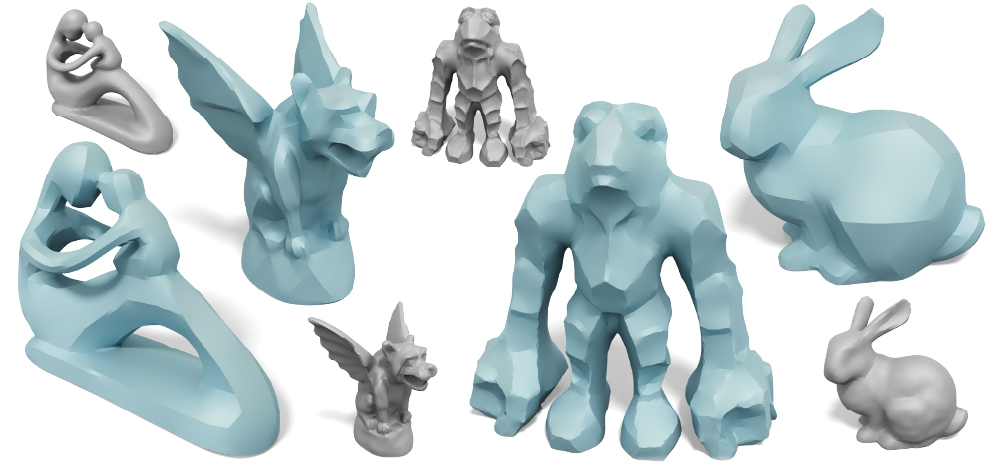}
  \vspace{-5pt}
  \caption{Our normal driven editing can be used to create many piece-wise developable surfaces (blue). Our method requires no remeshing and is fast enough for interactive modeling. \copyright cerberus333 (third) under CC BY-NC.} 
  \label{fig:developable}
  \vspace{-10pt}
\end{figure}
With our framework one can interactively create a variety of piece-wise developable shapes (see \reffig{developable}).
%
%
In \reffig{K_developable}, we evaluate our results by visualizing the discrete Gaussian curvature before and after running our developable flow. We can observe that the Gaussian curvature concentrates along the creases and results in a piece-wise developable surface. In the inset, we quantitatively demonstrate that our method effectively increases the developability of the mesh in \reffig{K_developable}.

%% file: sections/futureWork.tex
\section{Limitations \& Future Work}\label{sec:limitations}
Our method draws inspiration from Projective Dynamics \cite{BouazizMLKP14} to handle the case where target normals $\T$ are a function of output shape $\Vout$ (e.g., \reffig{update_t}, \reffignum{developable}). Although being fast and suitable for our intended interactive applications, it often struggles to converge to a highly accurate solution. 
Extending our optimization to, for example, Newton's method would be desirable for applications that desire highly accurate solutions.

\begin{figure}
  \centering
  \includegraphics[width=3.33in]{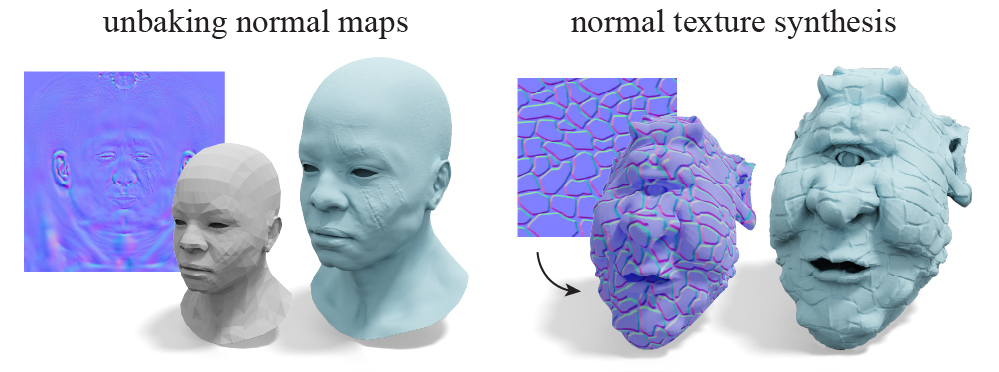}
  \vspace{-5pt}
  \caption{If one is interested in creating high-frequency geometric textures, we recommend to compute target normals via texture synthesis and then optimize the geometry via the normal-driven optimization. We demonstrate an example of ``unbaking'' normal maps (left) and an example of geometric texture synthesis (right).}
  \label{fig:synthesis}
  \vspace{-10pt}
\end{figure}

Our approach is restricted to a sphere as our reference shape $\sphere$, and uses the Gauss map to determine the correspondences between $\sphere$ and the input $\Vin$. As the Gauss map purely relies on surface normals to determine the map, the resulting map is ignorant to area distortion. 
This characteristic is beneficial to handle input shapes $\Vin$ that are very different (e.g., different genus) from a sphere because in these cases it is challenging to obtain a map with low area distortion. 
However, the price we have to pay is that we cannot support structured 
\begin{wrapfigure}[8]{r}{0.62in}
	\raggedleft 
	\vspace{-6pt}
	\hspace*{-0.7\columnsep}
	\includegraphics[width=0.75in, trim={4mm 0mm 1mm 0mm}]{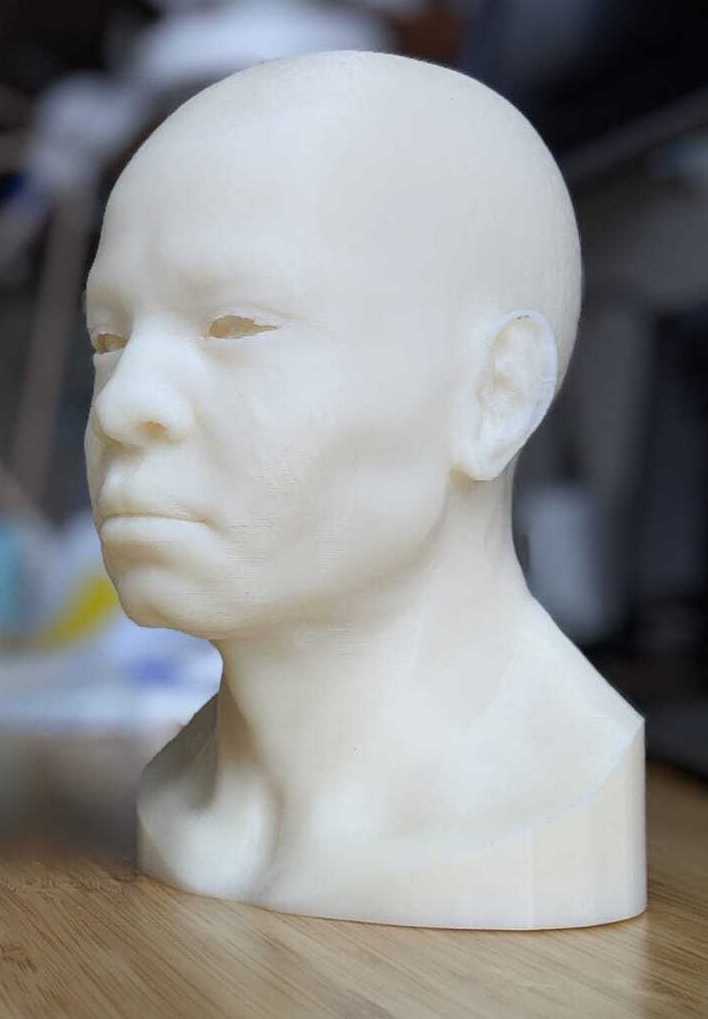} 
\end{wrapfigure} 
and high-frequency patterns (e.g., geometric texture synthesis). Thus, if one is interested in stylizing shapes with detailed textures, we suggest to first synthesize target normals on the surface directly \cite{WieLKT09} then perform the normal-driven optimization (\refsec{normalDriven_arap}). 
In \reffig{synthesis} we demonstrate this alternative by unbaking an existing normal map for manufacturing purposes (see the inset) and synthesizing normal textures from an image.

Our method currently supports manifold triangle meshes. 
Extending to non-manifold meshes, polygon meshes, volumetric meshes, and point clouds could be beneficial to handle real-world geometric data. 
Not every shape or normal capture sphere is valid to serve as the style shape of our algorithm. Discovering the validity of a style shape is important to understand the behavior of these novel modeling methods.
Removing the assumption about the source shape being a sphere could lead to a more general analogy-based shape editing. 
Based on the observation that surface normals are a promising geometric quantity to capture the style of a shape. Developing a better categorization of styles based on normals or exploring learning-based techniques on normals (instead of vertices) could lead to novel stylization methods.

\section*{Acknowledgements}
  Our research is funded in part by NSERC Discovery (RGPIN2017–05235, RGPAS–2017–507938), New Frontiers of Research Fund (NFRFE–201), the Ontario Early Research Award program, the Canada Research Chairs Program, the Fields Centre for Quantitative Analysis and Modelling and gifts by Adobe Systems, Autodesk and MESH Inc. 
  We thank Sheldon Andrews, Abhishek Madan, Silvia Sell{\'{a}}n, Oded Stein, Li-Yi Wei for helps on experiments. We thank members of Dynamic Graphics Project at the University of Toronto; Sarah Kushner, Abhishek Madan, Silvia Sell{\'{a}}n, Let{\'{i}}cia Mattos Da Silva, Towaki Takikawa for proofreading; John Hancock for the IT support. We thank all the artists for sharing a rich variety of 3D models.
  %

%% file: sections/appendix.tex
\appendix
\section{Local Step with $E_\arap$} \label{app:localStep}
Given a fixed $\U$, we obtain the optimal rotation for each vertex $k$ by solving the following minimization problem
\begin{align*}
  \R_k = \argmin_{\R_k \in\text{SO(3)}} \sum_{i,j \in \mathcal{N}_k} w_{ij} \| \R_k \dV_{ij} -  \dU_{ij} \|^2_2 + \lambda a_k\| \R_k \n_k - \tt_k \|_2^2
\end{align*}
The above optimization is an instance of the \emph{orthogonal Procrustes} which finds the best rotation matrix $\R_k$ to map a set of vectors ($\dV_{ij}, \n_k$) to another set of vectors ($\dU_{ij}, \tt_k$). We can re-write it into a more compact expression as:
\begin{align}\label{equ:localStep}
  &\R_k^\star = \argmax_{\R_k \in \text{SO}(3)} \ \Tr(\R_k \mX_k) \\
  &\mX_k = 
  \begin{bmatrix}
    \DV_k & \n_k
  \end{bmatrix}
  \begin{bmatrix}
    \mW_k & \vspace{3.5pt}\\
    & \lambda a_k
  \end{bmatrix}
  \begin{bmatrix}
    \DU_k^\top  \vspace{2pt}\\
    \tt_k^\top
  \end{bmatrix}.
\end{align}
where $\mW_k$ is a $|\mathcal{N}_k|$-by-$|\mathcal{N}_k|$ diagonal matrix of the cotangent weights $w_{ij}$, $\DV_k$ and $\DU_k$ are $3$-by-$|\mathcal{N}_k|$ matrices concatenating the edge vectors of the face one-ring at the rest and deformed states, respectively. 
One can then derive the optimal $\R_k$ from the SVD of $\mX_k = \mathcal{U}_k \sum_k \mathcal{V}_k^\top$
\begin{align}\label{equ:optRotation}
  \R_k = \mathcal{V}_k \mathcal{U}_k^\top,
\end{align}
up to changing the sign of the column of $\mathcal{U}_k$ so that $\det(\R_k ) > 0$.

\section{Global Step with $E_\arap$} \label{app:globalStep}
The global step updates the deformed vertex positions $\U$ from a fixed set of rotations $\R$ obtained via the local step. This boils down to solving the following problem 
\begin{align*}
  \U^\star = \argmin_{\U} \sum_{k \in V} &\sum_{i,j \in \mathcal{N}_k} w_{ij} \| \R_k \dV_{ij} -  \dU_{ij} \|^2_2
\end{align*}
We can expand this energy as
\begin{align*}
  &\sum_{k \in V} \sum_{i,j \in \mathcal{N}_k} w_{ij} \| \R_k \dV_{ij} -  \dU_{ij} \|^2_2 \\
  &\quad = \sum_{k \in V} \sum_{i,j \in f_k} w_{ij} \dU_{ij}^\top \dU_{ij} - 2 w_{ij} \dU_{ij}^\top \R_k \dV_{ij} + \text{constant} 
\end{align*}
It is often convenient to express the summation in terms of matrices. 
We introduce a directed incidence matrix $\mA_k$ with size $|V|$-by-$|\mathcal{N}_k|$ to  represent the edge vectors in $\mathcal{N}_k$ as $\V^\top \mA_k$, and we use $\mM_k$ to represent a $|\mathcal{N}_k|$-by-$|\mathcal{N}_k|$ diagonal matrix of the weights $w_{ij}$.  
%
Then we can re-write the energy in terms of matrices as
\begin{align}\label{equ:arapGlobalStep}
  &\sum_{k \in V} \Tr(\mM_k \mA_k^\top \U \U^\top \mA_k) - 2 \Tr(\mM_k \mA_k^\top \U \R_k \V^\top \mA_k ) \nonumber\\
  &= \sum_{k \in V} \Tr(\U^\top \mA_k\mM_k \mA_k^\top \U) - 2 \Tr(\R_k\V^\top\mA_k\mM_k \mA_k^\top \U) \nonumber\\
  &= \Tr\Big(\U^\top \Big(\textstyle\sum_k \mA_k\mM_k \mA_k^\top \Big) \U \Big) -2 \Tr\Big(\Big(\textstyle\sum_k \R_k\V^\top\mA_k\mM_k \mA_k^\top\Big) \U\Big) \nonumber\\
  &= \Tr(\U^\top \mQ \U) -2 \Tr(\R \mK \U),
\end{align}

where $\R = \{ \R_k \}$ is the concatenation of all the rotations, $\mQ$ is a $|V|$-by-$|V|$ symmetric matrix, and $\mK$ is a $|9V|$-by-$|3V|$ matrix stacking the constant terms which can be computed during the precomputation. We can then find the optimal $\U$ by solving a linear system
\begin{align*}
  \mQ \U = \mK^\top \R^\top
\end{align*}
As we know from \cite{SorkineA07}, $\mQ$ is the cotangent Laplacian \cite{PinkallP93}. We can pre-factorize $\mQ$ to speed up runtime performance. With these pieces in hand, we can minimize our energy \refequ{arapNormalDriven} by iteratively performing the local and the global steps (see \refalg{normalDriven}).

\section{Generalize to $E_\farap$ and $E_\acap$}\label{app:regularizations}
Changing the regularization from $E_\arap$ to the membrane-only regularization $E_\farap$ (\refequ{farap}) requires to re-define $\R$ on each face and change the set of edge vectors to the three edge vectors of a triangle. These changes would lead us to replace the $\DV_k, \DU_k$ in the local step \refequ{localStep} to the three edge vectors of a face, and $a_k$ to the face area. In the global step, one only needs to update the incidence matrix $\mA_k$ in \refequ{arapGlobalStep} to a $|V|$-by-$|f_k|$ matrix containing the three edge vectors information. 

Deploying the as-conformal-as-possible regularization $E_\acap$ (\refequ{acap}) changes the local step to solve an instance of the isotropic orthogonal Procrustes problem, where an analytical solution has been derived in \cite{schonemann1970fitting}. In short, one can obtain the optimal rotation the same way as \refequ{optRotation}, and compute the optimal scaling $s_k$ analytically as
\begin{align*}
  s_k = \frac{\Tr(\mW_k \DU^\top_k \R_k \DV_k) + \lambda a_k \n_k^\top \tt_k}{\Tr(\mW_k \DV^\top_k \DV_k) + \lambda a_k \n_k^\top \tt_k}.
\end{align*}
When assembling the matrices for the global step, using $E_\acap$ would require replacing $\R_k$ with $s_k\R_k$.

\section{Projective Dynamics for Dynamic Target Normals}\label{app:PD}
We draw inspiration from \emph{projective dynamics} \cite{BouazizMLKP14} to handle cases where the target normal $\T$ is a function of output geometry $\Vout$. Let us first define 
\begin{align*} 
  \T = \argmin E_N(\U)
\end{align*}
as a minimizer of an energy $E_N$ defined on the output shape. In our cases, $E_N$ could be the distance to the closet normals or the developable energy \cite{SteinGC18}. With this definition, we re-write \refequ{arapNormalDriven} as
\begin{align*} 
  &\min_{\U,\R}\ \sum_{k \in V} \sum_{i,j \in\mathcal{N}_k} w_{ij} \| \R_k \dV_{ij} -  \dU_{ij} \|^2_2  + \lambda a_k \| \R_k \n_k  - \tt_k \|_2^2,\\
  &\text{ subject to}\ \T = \argmin E_N(\U)
\end{align*}
This reformulation allows us to directly deploy the projective dynamics solver by first projecting $\T = \{ \tt_k\}$ to the ``constraint'' $E_N$, fixing $\tt_k$, and solving the original problem as \refequ{arapNormalDriven} via the local/global solver to get $\U$ at the next iteration. We then iterate this procedure (see \refalg{normalDriven}) until convergence. 
This expression enables us to plug-and-play different $E_N$ for different modeling objectives.

\section{Normal Driven Developable Surfaces} \label{app:developable}
Our normal-driven editing can be used to create developable surfaces by specifying a set of target normals that are developable.
Stein et al. \cite{SteinGC18} propose a characterization of discrete developability based on face normals of a vertex one-ring.
In short, if all the one-ring face normals correspond to a common plane or two planes, then this local one-ring is piecewise developable. 

With this characterization, we can easily get a set of ``developable'' face normals by (1) visiting all the one-ring faces of a vertex, (2) performing a small principle component analysis on the face normals for each one-ring, and (3) projecting the normals to one or two common planes by zeroing out the components correspond to the smallest eigenvalues. 
By using a different threshold to decide whether to zero out the smallest or the smallest two components, we can control the amount of creases in the developable approximation (see \reffig{control_developable}).
As each face will receive three (possibly) different developable normals from the previous procedure, we simply average them to get the target face normals. 
We perform this developable normal computation at each iteration in parallel, which corresponds to the Line 9 of \refalg{normalDriven}.